\documentclass[a4paper,11pt]{spieman}  
\usepackage{amsmath,amsfonts,amssymb}
\usepackage{graphicx}
\usepackage{setspace}
\usepackage{tocloft}
\usepackage{units}
\usepackage{upgreek}
\usepackage{bbm}
\usepackage{lineno}
\usepackage{textcomp}
\usepackage[absolute,overlay]{textpos}
\usepackage{helvet}

\title{Glass tube cutting with aberration-corrected non-diffracting ultrashort laser pulses}

\author[a]{Henning Rave}
\author[a]{Henning Heiming}
\author[a]{Patrick Szumny}
\author[a]{Myriam Kaiser}
\author[a]{Jonas Kleiner}
\author[a,*]{and \mbox{Daniel Flamm}}
\affil[a]{TRUMPF Laser- und Systemtechnik GmbH, Johann-Maus-Str.\,2, 71254 Ditzingen, Germany}

\cftpagenumbersoff{figure}
\cftpagenumbersoff{table} 
\begin{document}
\maketitle

\begin{abstract}
The separation of complex inner and outer contours of glass articles with curved surfaces using ultrashort pulsed lasers is reported. Single-pass, full-thickness modifications along the entire substrate are achieved using a processing optics that allows for beam shaping of non-diffracting beams and, additionally, for aberration compensation of phase distortions occurring at the curved interface. The glass articles finally separated by thermal stress or via selective etching meet the demands of the medical industry in terms of micro-debris, surface quality and processing speed.
\end{abstract}

\begin{spacing}{1}


{\noindent \footnotesize\textbf{*}Address all correspondence to Daniel Flamm: \linkable{daniel.flamm@trumpf.com}}


\section{Introduction}
\label{sect:intro} 
\begin{textblock*}{16cm}(2.375cm,1cm) 
   \centering
  \tiny \textsf{Henning Rave, Henning Heiming, Patrick Szumny, Myriam Kaiser, Jonas Kleiner, and Daniel Flamm "Glass tube cutting with aberration-corrected non-diffracting ultrashort laser pulses," Optical Engineering 60(6), 065105 (18 June 2021); \url{https://doi.org/10.1117/1.OE.60.6.065105}.}
\end{textblock*}
\begin{textblock*}{17.5cm}(1.8cm,27cm) 
   \centering \small 
   \textsf{
   © 2021 Society of Photo‑Optical Instrumentation Engineers (SPIE). One print or electronic copy may be made for personal use only. Systematic reproduction and distribution, duplication of any material in this publication for a fee or for commercial purposes, and modification of the contents of the publication are prohibited. \url{https://doi.org/10.1117/1.OE.60.6.065105}.}
\end{textblock*}
\noindent
With exceptional chemical, mechanical and thermal properties, glasses and glass ceramics are the materials of choice when vital substances need to be transported or long-term stored\cite{mahoney1991changes, knowles2006effects,wiwanitkit2006glass, el2011glasses}. The wide range of medical functionalities of ampoules, tubes, vials, syringes, etc.~requires equally diverse tools for precise and efficient processing. Established concepts known from laser machining of flat glass articles can be applied here only to a certain extent, since locally curved surfaces usually yield aberrated laser foci. This, in turn, prevents the controlled and efficient energy deposition into the volume of the glass substrate, as required, for example, when cutting transparent materials with elongated, non-diffracting focus distributions.\cite{Bhuyan2010, bhuyan2015high, marjanovic2017laser, Flamm2015,jenne2020faci}

We will present methods and processing optics enabling full-thickness, single-pass material modifications of glass tubes using ultrashort laser pulses. Here, sensitive phase correction is applied to generate adapted non-diffracting beams retaining their remarkable features even for strong curvatures (local radii $<\unit[5]{mm}$) or large wall thicknesses ($>\unit[1]{mm}$).\cite{rave2021glass} Resulting volume modifications allow the separation of outer contours with high edge qualities through the application of thermal or mechanical stress. Thermal separation strategies are being sought for transparent materials with low thermal expansion coefficients such as for borosilicate glasses in particular. The developed optics concept is still realized digital-holographically in the present experiments\cite{jenne2018,Jenne2018b, flamm2020structured}, however, we will also introduce options for the implementation as industrially suitable processing optics with stationary beam shaping components.\cite{Flamm2019, flamm2020structured} In fact, one of the concept's major benefits is the optical realization with standard components (axicons, cylindrical lenses).

The combination of discussed optical concepts with industrially well-established ultrashort-pulsed laser sources exhibiting pulse-specific control of pulse trains (bursts)\cite{herman2003burst} and adjustment of the repetition rate with single-pulse precision (pulse on demand)\cite{Jansen2018} enables fast ($>\unit[100]{mm/s}$) machining of outer and inner contours. Extracting complex internal contours is a challenge when processing both flat and curved glass articles, as no material is expelled during the laser modification process. We will therefore additionally introduce a selective laser etching strategy\cite{kaiser2019selective, flamm2020structured} enabling the processing of curved glass articles exhibiting vias of different dimensions and complex geometries. The glass articles fabricated this way meet the high quality criteria of the medical industry especially with regard to micro-debris and surface quality.

\section{Aberration corrected non-diffracting beams}\label{sec:optics}
\begin{figure}
    \centering
    \includegraphics[width=1\textwidth]{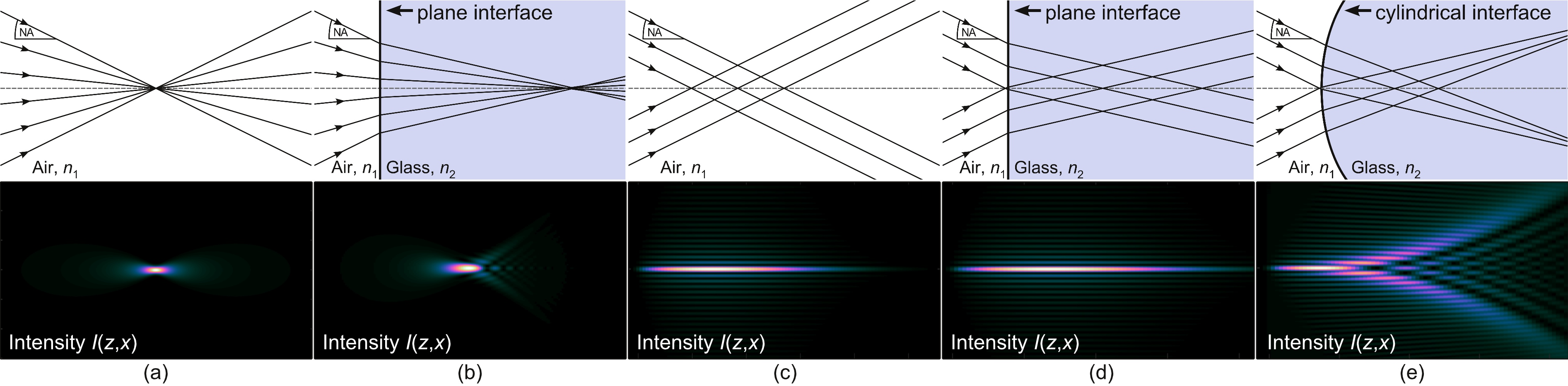}
    \caption{Ray optical and wave optical representation of non-aberrated and aberrated focus distributions. As for the wave optical case, normalized intensity cross sections $I\left(x,y=0,z\right)$ are shown. Clear length or intensity units are not labeled by intention, since a qualitative discussion is sufficient here. (a) Gaussian focus in air vs.~(b) spherically aberrated Gaussian focus focused behind a plane glass interface \cite{mauclair2008ultrafast, itoh2009spherical}. (c) Bessel-Gaussian focus in air vs.~(d) Bessel-Gaussian focus behind a plane glass interface. (e) Aberrated Bessel-Gaussian focus after propagation behind a cylindrical interface. Here and in some of the following figure we make use of D.~Green's color scheme representation\cite{Green2011}.}
    \label{fig:spher}
\end{figure}

Today, the use of non-diffracting beams for processing transparent materials is state-of-the-art.\cite{Kumkar2014, Flamm2019, jenne2020faci} Various processing optics based on refractive, reflective or diffractive axicons are available\cite{marjanovic2017laser, billaud2020high, PATkumkar2020optical} with benefits regarding focal length adjustment, working distance or crack control. The typically generated Bessel-Gaussian focus distribution exhibits several salient features useful for micro-machining, for a review see Ref.\,\citenum{Flamm2019}. Usually, particular emphasis is placed on self-healing. However, we would like to point out here on the ability to resist spherical aberrations which usually appear when light is focused deep behind an optical interface\cite{mauclair2008ultrafast, itoh2009spherical}. Depending on refractive index differences and used numerical apertures (NAs) the resulting focal field exhibits a tremendous loss of peak intensity as well as increased transverse and longitudinal dimensions\cite{itoh2009spherical, Bergner2018}, see comparison in Fig.\,\ref{fig:spher}\,(a) and (b). To get an impression about the interface's impact for this focusing situation ($\text{NA} \sim\unit[0.4]{}$, Gaussian pupil illumination, ideal paraxial focusing, $\Delta n = 0.5$, $\lambda = \unit[1]{\upmu m}$, focal depth behind interface $\sim\unit[1]{mm}$), the beam quality in terms of beam propagation ratio\cite{siegman1998maybe} may degrade here by several factors from the diffraction limit ($M^2_{\text{real}}\gtrsim5$).\cite{Tillkorn2018} Equally significantly reduced would be the corresponding Gaussian Strehl-ratio\cite{mahajan2005strehl} [$\text{SR}_{\text{real}}\lesssim0.6$, for the particular case shown in Fig.\,\ref{fig:spher}\,(b)]. The spherical aberration effect discussed frequently occurs in this or a similar form during, e.g., microscopy and micromachining, so that compensation is often required, e.g.~by applying index matching\cite{gibson1991experimental} or using liquid crystal-based spatial light modulators \cite{mauclair2008ultrafast, itoh2009spherical}.

\begin{figure}
    \centering
    \includegraphics[width=0.9\textwidth]{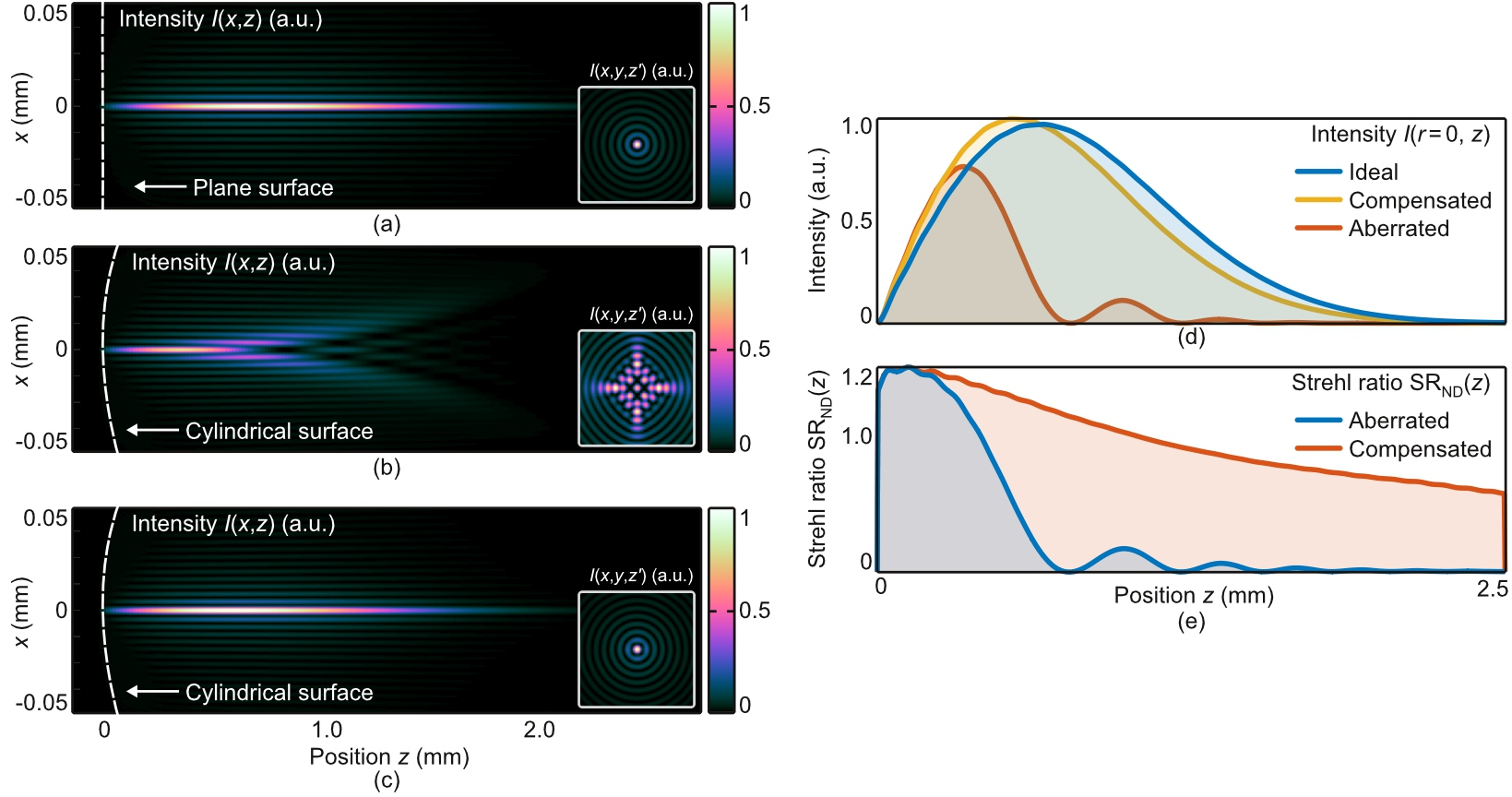}
    \caption{(a) Simulated longitudinal intensity profiles $I\left(x,y=0,z\right)$ of Bessel-Gaussian beams behind a plane and (b) a cylindrical surface. The insets depict transverse intensity distributions at a propagation distance of $\sim\unit[1]{mm}$ behind the surface $I\left(x,y,z=\unit[1]{mm}\right)$. Additionally, (c) intensity profiles are depicted where aberration compensation is applied. (d) Corresponding plots of the on-axis intensities $I\left(r=0,z\right)$ of the three cases. (e) The deduced Strehl-ratio $\text{SR}_{\text{ND}}$ for the aberrated and aberration-compensated situation are shown in. The fact that $\text{SR}_{\text{ND}}$ partially exceeds values of $1$, see (e), is caused by the cylindrical surface which exhibits additional focusing properties resulting in slightly higher intensities for $I_{\text{real}}\left(r=0, z\right)$ especially for low $z$ values.}  \label{fig:correct}
\end{figure}

A comparable focus situation, but now for a Bessel-Gaussian beam is depicted in  Fig.\,\ref{fig:spher}\,(c) and (d). Different to the Gaussian case, where a broad spatial frequency spectrum is generating the focus, there is only a single radial component of the wavevector $k_r$ shaping the characteristic elongated interference pattern \cite{McGloin2005}. Since the same refraction condition applies to all of these field components at the interface, the resulting Bessel-Gaussian beam remains undisturbed. Merely the beam length scales with the present refractive index. This ``natural'' resistance to spherical aberrations [Meant here is the situation in which a non-diffracting beam can be focused vertically behind a plane interface to any depth and with any high NA, without the otherwise usual spherical aberrations occurring (as for initial plane or Gaussian raw beams)\cite{itoh2009spherical, Bergner2018a}. A spherical phase aberration, e.g., in terms of Zernike wavefront distortions will yield perturbed non-diffracting focus distributions, too. \cite{merx2019fast} It should also be pointed out at this point that the effect of a phase aberration to a non-diffracting beam depends strongly on where it is imposed (near-field, far-field or in between). Considering the aberrated non-diffracting beams depicted, e.g., in Figs.\,\ref{fig:spher}\,(e) or \ref{fig:correct}\,(b), typical near-field aberrations are induced as the starting point of the focus distribution equals the position of aberration surface.] enables to modify transparent materials with thicknesses $>\unit[10]{mm}$ in a single pass.\cite{Flamm2019}

A glass tube characterized by a cylindrical interface of radius $R_{\text{T}}$ represents the simplest form of a curved glass surface. As can be seen in the schematic of Fig.\,\ref{fig:spher}\,(e), this interface acts as cylindrical lens resulting in an aberrated Bessel-Gaussian focus distribution.\cite{rave2021glass} Depending on the glass tube geometry (refractive index $n_2$, radius of curvature $R_{\text{T}}$ and wall thickness $w_{\text{T}}$) the ideal on-axis constructive interference is disturbed in a way that strong intensity modulations appear. Resulting heterogeneous material modifications will prevent obtaining high quality glass edges or will, basically, prevent the separation along a desired contour. To evaluate the quality of the resulting perturbed foci we follow the suggestion of Merx \textit{et al.}\cite{merx2019fast} and compute the Strehl-ratio of non-diffracting beams $\text{SR}_{\text{ND}}$ by comparing ``real'' (thus, potentially perturbed) on-axis intensities ($r=0$) with corresponding ideal cases. This extended Strehl-ratio is, thus, no longer a single representative parameter (see discussion in this section's first paragraph on spherically aberrated Gaussian foci) but evaluates the focus quality along the elongated focus zone in propagation direction $z$
\begin{gather}
    \text{SR}_{\text{ND}}\left(z\right) = \frac{I_{\text{real}}\left(r=0, z\right)}{I_{\text{ideal}}\left(r=0, z\right)}.
\end{gather}
Simulated longitudinal $I\left(x,y=0,z\right)$ and transverse $I\left(x,y,z=\unit[1]{mm}\right)$ intensity  profiles of non-diffracting beams (Bessel-Gaussian beams) focused behind plane and cylindrical interfaces, respectively, are depicted in Fig.\,\ref{fig:correct}\,(a) and (b). The cylindrically curved surface causes characteristic deviations from the ideal focus shape, see e.g., clear on-axis modulations or the rhombus-like transverse intensity profile.
This is confirmed by the on-axis intensity plots $I\left(r=0,z\right)$ shown in Fig.\,\ref{fig:correct}\,(d) and the deduced Strehl-ratio $\text{SR}_{\text{ND}}\left(z\right)$ in Fig.\,\ref{fig:correct}\,(e). In the aberrated case, a clear minimum at $z\approx\unit[0.8]{mm}$ is evident for $\text{SR}_{\text{ND}}$ which will lead to insufficient or altered material degradations at these points and glass tubes with wall thicknesses of $w_{\text{T}}\gtrsim\unit[0.8]{mm}$ will be difficult to separate.

\begin{figure*}
    \centering
    \includegraphics[width=1\textwidth]{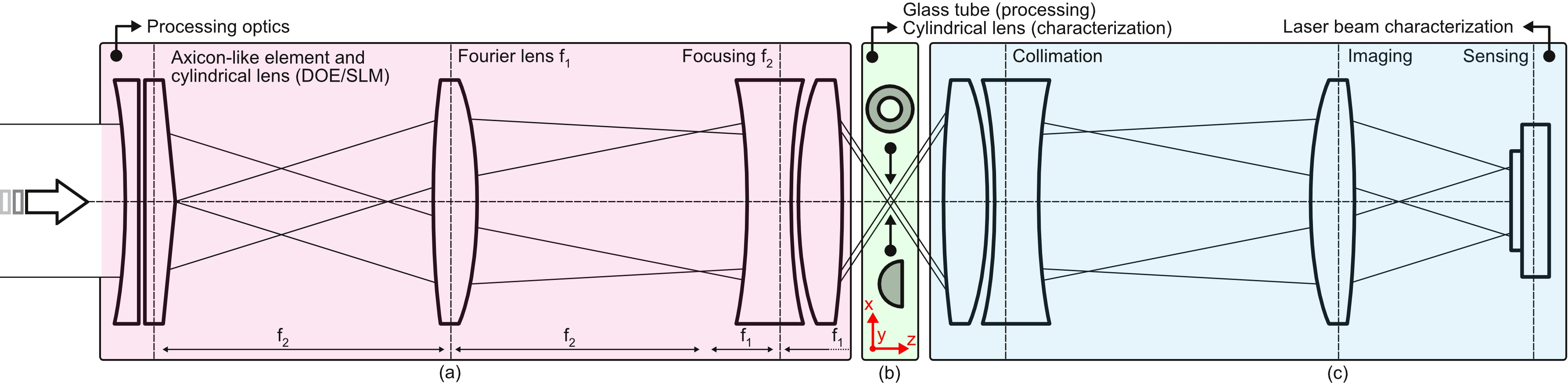}
    \caption{Optical setup for processing of glass tubes and for laser beam characterization. (a) The processing optics fed by free-space or fiber-guided laser pulses consists of the axicon-like element for beam shaping and aberration compensation (for example realized with cylindrical lens and axicon) and a telescopic setup for imaging the non-diffracting beam into the glass tube (b). (c) Laser beam characterization is realized by the inverted optical setup. A telescope is imaging the focus distribution behind a plano-cylindrical lens of adapted radius, see (b), onto the camera for caustic measurements.}
    \label{fig:sschool}
\end{figure*}

To correct the aberrations induced, we assume that the cylindrical glass surface acts as ideal thin element with quadratic phase modulation of a cylindrical lens\cite{goodman2005introduction, saleh2019fundamentals}
\begin{gather}
\Phi_{\text{cyl}}'\left(x\right) = \uppi x^2 / \lambda f_{\text{cyl}}',
\end{gather}
with the lens' focal length $ f_{\text{cyl}}' \approx  R_{\text{T}} / \Delta n$. The amplitude of the incident optical field remains unaffected by the thin surface. Optical access to the $z$-position of the curved surface is provided virtually using a telescopic setup with given magnification $M=f_2/f_1$, cf.~Fig.\,\ref{fig:sschool}\,(a).
If the start ($z$-position) of the non-diffracting beam is close to the curved optical interface, the telescopic setup allows to perform the beam shaping and the aberration correction within a single element, see axicon-like element depicted in Fig.\,\ref{fig:sschool}. In the present case, the magnified phase aberration compensation $\Phi_{\text{cyl}} \approx -M^2\Phi_{\text{cyl}}'$ may be realized as refractive element with a cylindrical surface of radius $R_{\text{cyl}} \approx -M^2R_{\text{T}}$ and focal length $f_{\text{cyl}} \approx -M^2f_{\text{cyl}}'$, respectively (for example: $R_{\text{T}}=\unit[5]{mm}$, $M^2=400$ and $\Delta n = 0.5$ results in a standard cylindrical lens with $f_{\text{cyl}} = -\unit[4]{m}$). Equally conceivable is the diffractive realization as single element with muliplexed phase modulations of an axicon $\Phi_{\text{axi}}$\cite{chen2019generalized, jenne2020faci} and the described cylindrical lens $\Phi_{\text{cyl}}$. We prefer the latter variant because it eliminates the need for laborious adjustment of the two components (axicon with respect to cylindrical lens). Thus, the generation of the non-diffracting beam and its aberration correction is performed by a single central beam shaping element\cite{Flamm2019}. For the experiments conducted in this work the phase of the raw beam was modulated using a flexible liquid-crystal-on-silicon based spatial light modulator (SLM).
It is planned later to equip a processing optics with a stationary diffractive optical element with clear advantages regarding power/energy suitability and zero-order power management.\cite{Flamm2019}

\section{Single-pass, full-thickness modifications of glass tubes}\label{sec:exp}
We use the introduced optical concepts from previous section, cf.\,Fig.\,\ref{fig:sschool}, to directly measure the intensity distribution of the non-diffracting beam behind the curved glass surface. For this, the plano cylindrical lens [Fig.\,\ref{fig:sschool}\,(b)] is aligned into the setup with respect to the starting point and optical axis of the non-diffracting focus. The telescopic setup from the laser beam characterization part of our experimental setup, see Fig.\,\ref{fig:sschool}\,(c), is imaging the intensity distribution inside the cylindrical lens with $R_{\text{T}} = \unit[4]{mm}$ onto the camera being moved through the magnified focus with a $z$-axis system.
\begin{figure}
    \centering
    \includegraphics[width=0.9\textwidth]{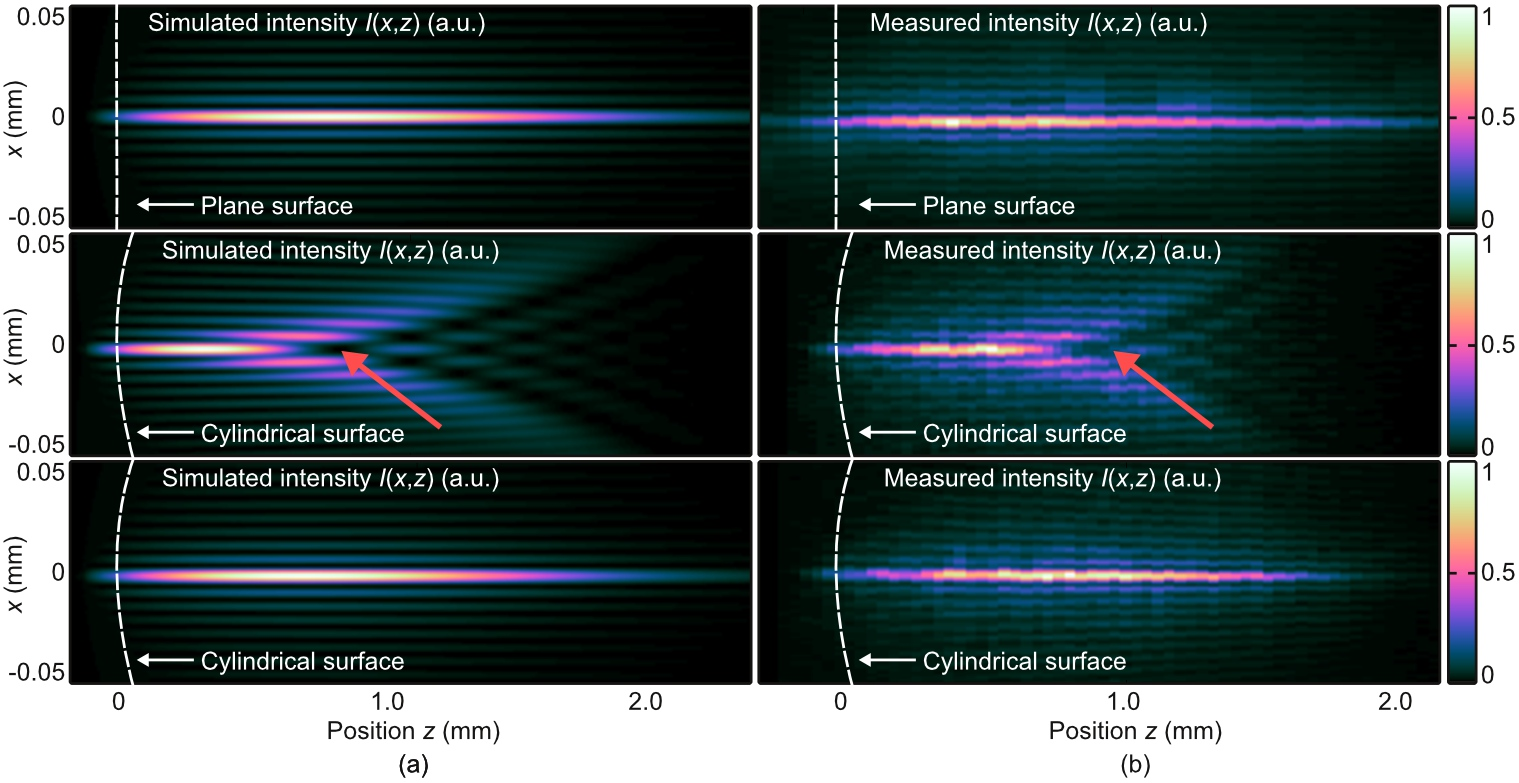}
    \caption{(a) Simulated and (b) measured intensity cross sections $I\left(x,z\right)$ of non-diffracting beams propagating behind a plane interface (top row), a cylindrically curved interface without aberration correction (second row) and with applied aberration correction (third row). Please note, the excellent agreement of distinct intensity features such as on-axis intensity modulations, see red arrows in second row.}
    \label{fig:meas01}
\end{figure}
Results of different caustic measurements are depicted in Fig.\,\ref{fig:meas01}. For completeness and for direct comparison, corresponding simulations [Fig.\,\ref{fig:meas01}\,(a)] are placed ahead of the experimental findings shown in Fig.\,\ref{fig:meas01}\,(b) as intensity cross sections $I\left(x,z\right)$. In the first row, the ``ideal'' case of focused non-diffracting beams behind plane interfaces is depicted. In the second row, simulated and measured cross sections behind curved surfaces can be seen. Finally, in the last row, the aberration correction is applied by additionally displaying a holographic cylindrical lens transmission with the SLM. All three measured cases convince with high agreements to the respective simulations and confirm our theory. Considering distinguished features, such as the intensity modulation on the optical axis $I\left(r=0, z\right)$ in the aberrated case (second row, see red arrows), the distributions also agree in details. It can be clearly seen that in the second case the on-axis intensity drops almost to $0$ at $z\approx \unit[0.8]{mm}$. A successful energy deposition at these $z$-positions is prevented due to the destructive interference which cannot be compensated by increasing pulse energies either. With applied aberration correction (third row), elongated, high on-axis intensities can be achieved almost up to $z\approx \unit[2]{mm}$ in this particular case.\cite{rave2021glass}

\begin{figure}
    \centering
    \includegraphics[width=0.9\textwidth]{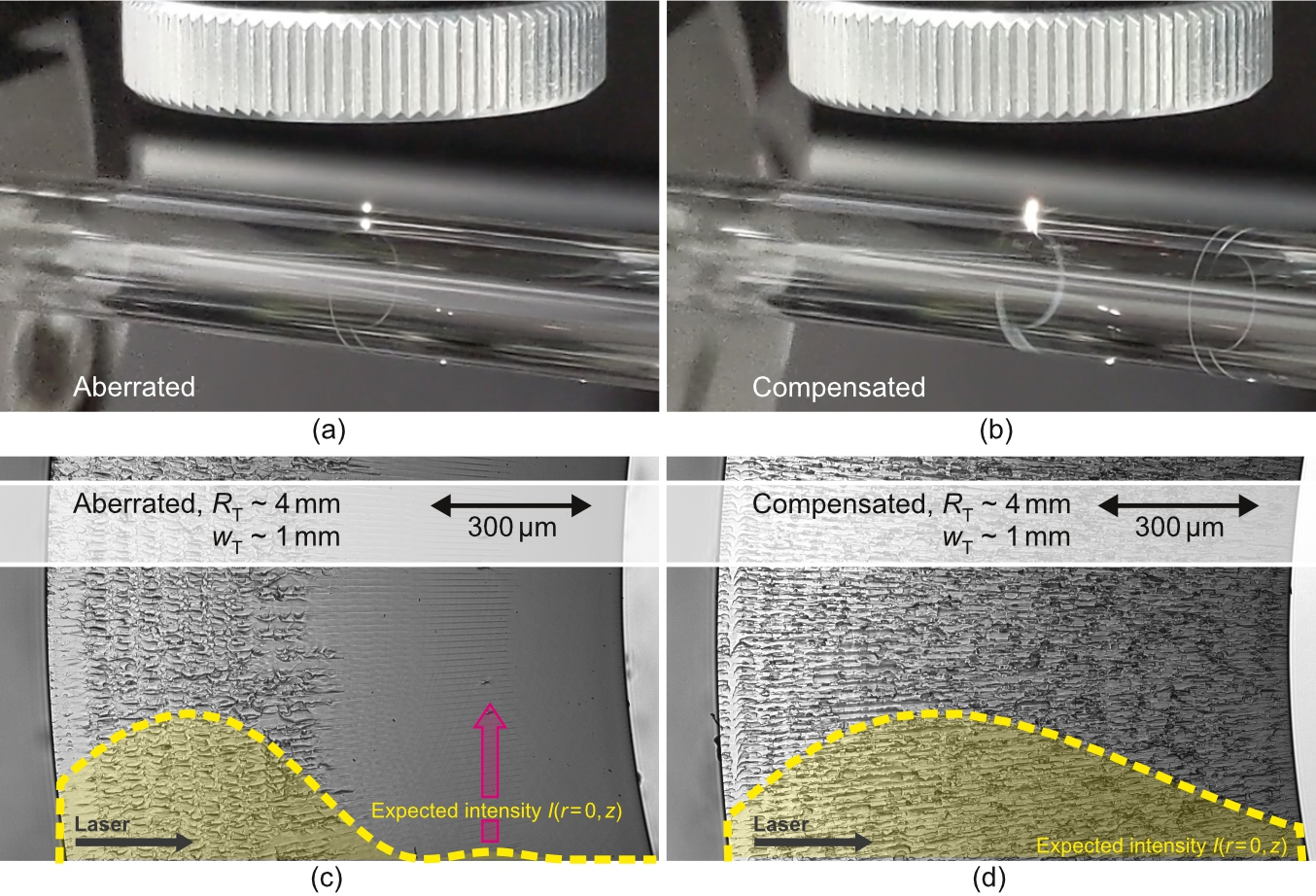}
    \caption{Recording of the laser modification process and microscope images of borosilicate tube edges after mechanical separation. Plasma response of the glass tube, camera-recorded during inducing laser modification (a) without and (b) with applying the phase aberration correction. The borosilicate tube is rotated along its cylindrical axis with respect to the optical axis of the processing optics illuminating the tangential surface perpendicularly from the top, see housing of the focusing unit. Differences in the length of both the plasma glow and the volume modifications are visible to the naked eye, see also Media 1. The entirety of modifications acts as breaking layer at which the actual separation is achieved mechanically, in this particular case. (c) Elongated type III modifications \cite{Itoh2006} extend from the curved entrance surface (left) to approximate half the wall thickness in case of processing with an aberrated non-diffracting beam. (d) The processing with aberration-corrected non-diffracting beams, on the other hand, allows to generate uniform modifications along the entire wall thickness. The yellow graph indicates the on-axis intensity distribution in both cases, mainly responsible for the resulting modifications. Please note, that our simulations even predict the position of weak modifications caused by the on-axis intensity's side maximum, see purple arrow in (a).}
    \label{fig:roar-Seite001.jpeg}
\end{figure}

In the following, material processing experiments are conducted using the processing optics part of our setup, cf.\,Fig.\,\ref{fig:sschool}\,(a). The cylindrical lens used for above discussed laser beam characterization experiments is replaced by glass tubes of different geometries $\left[R_{\text{T}} = \left(1 \dots 20\right)\,\text{mm}, w_{\text{T}} = \left(0.1 \dots 2\right)\,\text{mm}\right]$. Ultrashort laser pulses at $\lambda = \unit[1030]{nm}$ are illuminating the axicon-like element shaping the aberration-corrected non-diffracting beam. The optical axis of the processing optics is aligned perpendicular to the glass tube's tangential surface, see Fig.\,\ref{fig:roar-Seite001.jpeg}\,(a) and (b). During the laser modification step, tubes are rotated around their symmetry axis, see Media 1, enabling feed rates of $\gtrsim\unit[100]{mm/s}$ by employing $\unit[20]{W}$ of average power from a TruMicro 2030 laser system. Pulse energy, burst parameters and spatial pulse distance are adequately chosen to (mainly) generate elongated type III modifications\cite{Itoh2006} along the entire wall thickness of borosilicate tubes with a single pass. The entirety of modifications acts as breaking layer at which the actual separation is achieved. Full-thickness modifications are confirmed by investigating the modification quality of the outer contour's edge.
In Fig.\,\ref{fig:roar-Seite001.jpeg}\,(c) and (d) microscope images of details of tube edges are depicted that were separated mechanically. Elongated modifications ranging from the curved entrance surface to approximately half the wall thickness can be seen for the case of processing with an aberrated non-diffracting beam, see Fig.\,\ref{fig:roar-Seite001.jpeg}\,(c). In contrast, the processing with aberration corrected non-diffracting beams enables the formation of uniform modifications along the entire thickness, as shown in Fig.\,\ref{fig:roar-Seite001.jpeg}\,(d). Although despite insufficiently long modifications, mechanical separation was successful in the first case [Fig.\,\ref{fig:roar-Seite001.jpeg}\,(c)], the required breaking force was increased by several factors compared to the case with full-thickness modifications [Fig.\,\ref{fig:roar-Seite001.jpeg}\,(d)]. Additionally, this increases the probability of generating chippings and a higher level of micro debris during the separation process.
\begin{figure}
    \centering
    \includegraphics[width=1\textwidth]{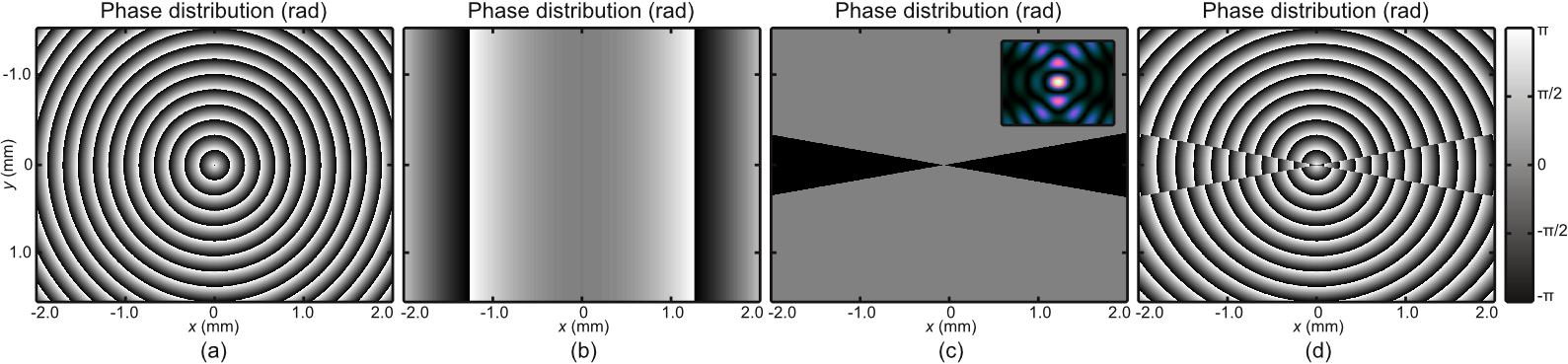}
    \caption{Central details of phase modulations required for beam shaping and aberration correction in modulo-$2\uppi$ representation. (a) Holographic axicon carrier (equivalent to the phase modulation of an ideal thin $\sim 1^{\circ}$-axicon)\cite{Leach2006}. (b) Phase modulation of a concave cylindrical lens (equivalent to a thin refractive version with focal length of $f_{\text{cyl}} \approx \unit[-800]{mm}$).\cite{saleh2019fundamentals} (c) Azimuthal phase dependency to induce a preferential direction to the on-axis intensity maximum (see inset).\cite{chen2019generalized, jenne2020faci} (d) Final phase mask as superposition of the three aforementioned ones [(a)--(c)] displayed by a flexible spatial light modulator or realized by a static diffractive optical element.\cite{flamm2020structured}}
    \label{fig:phama}
\end{figure}

Various groups currently report on techniques to simplify the separation of laser-modified sheet-like glass articles. This involves the controlled generation of cracks between the elongated modifications, whose orientation is aligned with the processing direction.\cite{hendricks2016, Meyer2017, Dudutis2018} For this purpose, elliptical voids are generated using asymmetric non-diffracting beams. If mechanical stress is applied to the modified glass, the stress accumulates at the small radii of the elliptical defects resulting in crack discharges along the long main axes.\cite{jenne2020faci} With this process, crack-connected modifications have already been demonstrated that were more than $\unit[100]{\upmu m}$ apart.\cite{flamm2020structured} The required asymmetric non-diffracting beams can be generated in various ways, e.g, by inducing near-field aberrations\cite{Dudutis2018} or by applying far-field amplitude filter\cite{Meyer2017}. To facilitate glass tube separation, elliptical non-diffracting beams are also employed, generated here via so called generalized axicons\cite{chen2019generalized}. In addition to the constant radial phase slope known from conventional axicons\cite{McGloin2005}, an azimuthal distribution is imprinted to the illuminating optical field.\cite{flamm2020generalized} This allows the generation of non-diffracting beams with tailored transverse intensity distributions including elliptical on-axis intensity maxima, for details see Chen \textit{et al.}\cite{chen2019generalized}. Due to symmetry breaking, these components are preferably realized diffractive optically via laser lithography\cite{flamm2020structured}. This offers the advantage that several optical functionalities can be combined in a single central beam shaping element. Figure \ref{fig:phama} depicts the phase transmission of the axicon-like element, cf. Fig.\,\ref{fig:sschool}\,(a), used for glass tube processing. It consists of the holographic axicon carrier\cite{Leach2006} 
[Fig.\,\ref{fig:phama}\,(a)] to form the non-diffracting beam, the aberration correction with a cylindrical lens [Fig.\,\ref{fig:phama}\,(b)] and an azimuthal $\uppi$-phase jump to break the radial symmetry (``generalized axicon'') [Fig.\,\ref{fig:phama}\,(c)]. The final phase mask being a superposition of the aforementioned ones, is shown in Fig.\,\ref{fig:phama}\,(d). For our fundamental experiments a liquid-crystal-on-silicon-based spatial light modulator is displaying this phase-only transmission function and acts as holographic axicon-like element, cf.~Fig.\,\ref{fig:sschool}\,(a). For reasons of power/energy resistance, a stationary diffractive solution in a processing optics is aimed\cite{flamm2020structured} with clear advantages regarding adjustment and compactness.

\section{Selected processing examples}
As already mentioned in Sec.\,\ref{sect:intro}, due to remarkable thermal properties, tubes made of borosilicate glass are of particular interest for the medical industry. We therefore focus on this material and would like to emphasize that the parameters presented in the following are strongly dependent on the respective glass type and associated tube geometry $\left(R_{\text{T}}, w_{\text{T}}\right)$.

Using the processing strategy explained in Sec.\,\ref{sec:exp}, borosilicate tubes of radius $R_{\text{T}} = \unit[3.9]{mm}$ and wall-thickness $w_{\text{T}} = \unit[1.05]{mm}$ were modified using aberration-corrected asymmetric non-diffracting beams. By exploiting $\unit[20]{W}$ average power from a TruMicro Series 2000 laser maximum feed rates of $\gtrsim \unit[100]{mm/s}$ ($\sim \unit[4]{Hz}$) were achieved. The pulse energy required was $\unit[240]{\upmu J}$ equally distributed to 4 subpulses with $\unit[3]{ps}$ pulse duration. The actual separation was realized mechanically and quantified using a four-point flexural load test.
\begin{figure}
    \centering
    \includegraphics[width=0.85\textwidth]{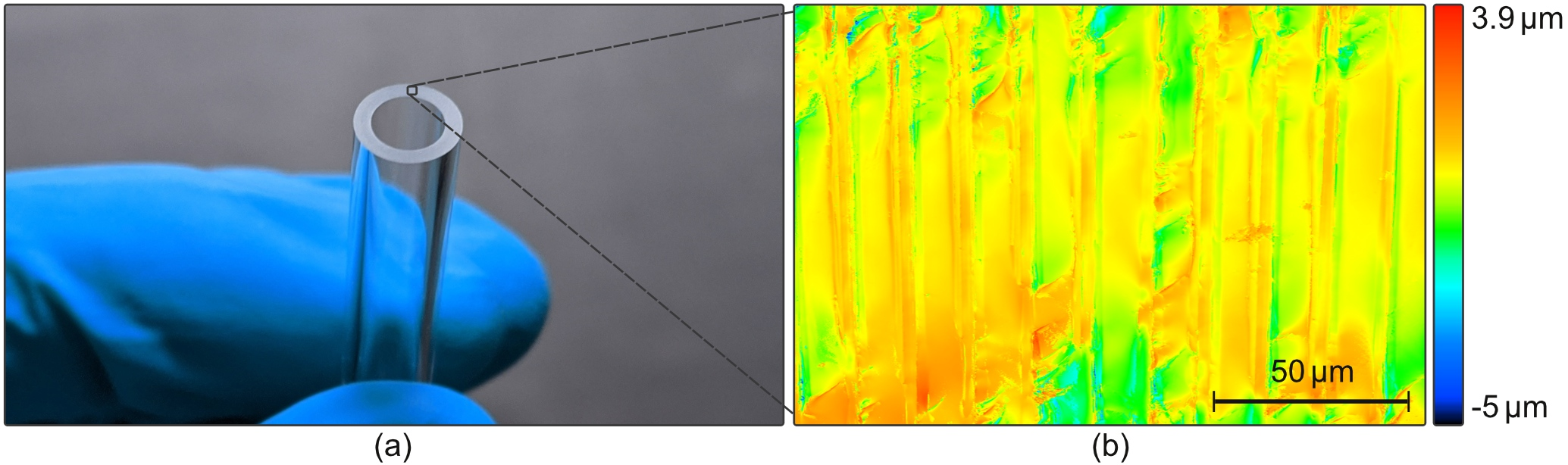}
    \caption{(a) Mechanically separated glass tube made of borosilicate glass (Radius $R_{\text{T}} = \unit[3.9]{mm}$, wall thickness $w_{\text{T}} = \unit[1.05]{mm}$). The reader may make a visual comparison with an unmachined tube edge as it is created from the melt after solidification. This is shown on each subfigure's right hand side of Media 1. Volume modifications were achieved using an adapted TOP Cleave cutting optics and a TruMicro Series 2000 laser in a single pass. (b) Mean surface roughness of the tube edge was determined to $S_a \lesssim \unit[1]{\upmu m}$ (area evaluation). By exploiting $\unit[20]{W}$ of average power, feed rates of $\gtrsim \unit[100]{mm/s}$ were achieved. Although not demonstrated here, further scaling to m/s-feed rates are feasible using industry-grade multi millijoule- and multi hundred-watt-class\cite{sutter2019high} ultrashort pulsed laser systems, such as a TruMicro Series 6000 laser.}
    \label{fig:exampleBlau}
\end{figure}
Figure \ref{fig:exampleBlau}\,(a) depicts the edge of a separated glass tube whose mean surface roughness was evaluated on the area to $S_a \lesssim \unit[1]{\upmu m}$ using a laser scanning microscope, see Fig.\,\ref{fig:exampleBlau}\,(b). A conventional microscope image of the tube edge with a similar geometry is already provided in Fig.\,\ref{fig:roar-Seite001.jpeg}\,(b). This laser processing result [Fig.\,\ref{fig:exampleBlau}\,(a)] can be compared with that of an unmachined tube edge as it is created from the melt after solidification. This is shown on each subfigure's right hand side of Media 1. The required flexural load for mechanical separation based on a four-point bending test was $\sim \unit[\left(50\pm20\right)]{N}$ (strong $R_{\text{T}}$-dependent). This comparatively low value, which was achieved in particular through the use of an aligned asymmetric non-diffracting beam and crack control, cf.~Fig.\,\ref{fig:phama}, enables a thermally induced mechanical stress-based separation for borosilicate tubes. Please note, that the stated laser parameters of this study (pulse energy and duration, pulse distance, burst modus, etc.) represent useful values and can form the basis for future investigations. However, we do not claim to have found the optimum. Depending on tube geometry and material adapted laser parameters need to be found. In addition, there will be a dependency on which separation process (for example: thermal vs.~mechanical) is actually aimed at.\cite{rave2021glass}

Up to this point, main focus was on outer contour separation of tubes made of transparent materials. A stationary processing optics was focusing adapted non-diffracting beams into the walls of tubes being rotated around their cylinder axis, cf.~Fig.\,\ref{fig:exampleBlau}\,(a) and (b). If, in addition to rotation about this cylinder axis [$y$-axis, see red coordinate system in Fig.\,\ref{fig:sschool}\,(b)], translation is also possible, modifications along complex inner contours are realized, see Media 2. However, a damage-free release is still challenging, since no material is expelled during the laser modification process. Established procedures for through glass vias fabrication from sheet-like substrates can be used here as well. One way to accomplish component extraction is to take advantage of the increased etchability of laser modified contours.
\begin{figure}
    \centering
    \includegraphics[width=1.0\textwidth]{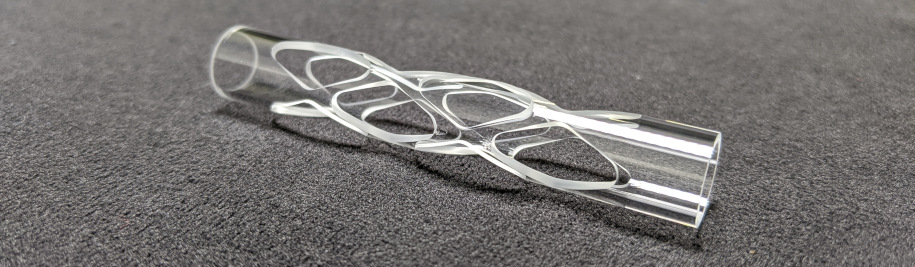}
    \caption{Separation of complex inner contours from borosilicate tubes (Radius $R_{\text{T}} = \unit[5]{mm}$, wall thickness $w_{\text{T}} = \unit[1]{mm}$) using a selective laser etching strategy (ultrasonic bath-assisted etching in $30$\,wt.-$\%$ KOH solution at $\unit[80^{\circ}]{C}$).\cite{kaiser2019selective} Laser modification step is visualized in Media 2.}
    \label{fig:edge}
\end{figure}
The process known as selective laser etching\cite{kaiser2019selective, flamm2020structured} can be applied to curved substrates, too, as proven by the sample depicted in Fig.\,\ref{fig:edge} where several vias (smallest contour radii below $\unit[1]{mm}$) were extracted from a borosilicate tube (radius $R_{\text{T}} = \unit[5]{mm}$, wall thickness $w_{\text{T}} = \unit[1]{mm}$). Here, ultrasonic bath-assisted selective etching was applied in a $30$\,wt.-$\%$ KOH solution at $\unit[80^{\circ}]{C}$.\cite{kaiser2019selective}

\section{Conclusion}
A concept for ultrashort pulsed laser cutting of glass tubes was developed and applied to borosilicate samples exhibiting radii of $R_{\text{T}} > \unit[1]{mm}$. The presented processing optics generates an aberration corrected non-diffracting beam which enables single-pass, full-thickness modifications with feed rates $\gtrsim \unit[100]{mm/s}$ using $\unit[20]{W}$ of average power from a TruMicro 2000 Series laser. Achieved edge qualities with mean surface roughness parameters of $S_a \lesssim \unit[1]{\upmu m}$ meet the demands of the medical industry regarding chipping and micro debris. The concept enables a thermal seperation strategy even for borosilicate samples proven by minor loads required for mechanical separation with flexural bend tests. As an outlook, the extraction of complex inner contours from glass tubes by means of a selective etching strategy was demonstrated.

\subsection*{Disclosures}
The authors declare no conflicts of interest.


\footnotesize

\vspace{2ex}
\noindent Biographies and photographs of the authors are not available.


\end{spacing}
\end{document}